\documentclass[twocolumn, tighten]{aastex63}
\hypersetup{linkcolor=red,citecolor=blue,filecolor=cyan,urlcolor=magenta}
\usepackage{xspace}
\usepackage{multirow}

\submitjournal{ApJ}
\shorttitle{Compton Hump Reverberation}
\shortauthors{Zoghbi et al.}
\graphicspath{{./}{figures/}}

\newcommand{\FeK}{\mbox{Fe K$\alpha$}\xspace}
\newcommand{\xmm}{{\it XMM-Newton}\xspace}
\newcommand{\nustar}{{\it NuSTAR}\xspace}

\newcommand{\swlong}{\mbox{SWIFT J2127.4+5654}\xspace}
\newcommand{\sw}{\mbox{SW J2127}\xspace}
\newcommand{\mcg}{\mbox{MCG--5-23-16}\xspace}

\defcitealias{2017ApJ...836....2Z}{Z17}
\defcitealias{2014ApJ...789...56Z}{Z14}

\begin{document}

\title{A Hard Look At Relativistic Reverberation in MCG--5-23-16 \& SWIFT J2127.4+5654: Testing the Lamp-Post Model}

\email{abzoghbi@umich.edu}
\author[0000-0002-0572-9613]{A. Zoghbi}
\affil{Department of Astronomy, University of Michigan, Ann Arbor, MI 48109, USA}

\author{J. M. Miller}
\affil{Department of Astronomy, University of Michigan, Ann Arbor, MI 48109, USA}

\author{E. Cackett}
\affiliation{Department of Physics, Wayne State University, Detroit, MI 48201, USA}

\begin{abstract}
X-ray reverberation mapping has emerged as a new tool to probe accretion in AGN, providing a potentially powerful probe of accretion at the black hole scale. The lags, along with relativistic spectral signatures are often interpreted in light of the lamp-post model. Focusing specifically on testing the prediction of the relativistic reverberation model, we have targeted several of the brightest Seyfert Galaxies in X-rays with different observing programs. Here, we report the results from two large campaigns with \nustar targeting \mcg and \swlong to test the model predictions in the 3--50 keV band. These are two of three sources that showed indications of a delayed Compton hump in early data. With triple the previously analyzed exposures, we find no evidence for relativistic reverberation in \mcg, and the energy-dependent lags are consistent with a log-linear continuum. In \swlong, although a continuum-only model explains the data, the relativistic reverberation model provides a significant improvement to the energy and frequency-dependent lags, but with parameters that are not consistent with the time-averaged spectrum. This adds to mounting evidence showing that the lag data is not consistent with a static lamp-post model. 

\end{abstract}


\section{Introduction} \label{sec:intro}
X-ray radiation is unique in its ability to probe the immediate vicinity of accreting black holes. Spectral signatures in the form of distorted and broadened emission lines have been observed for decades \citep{1995Natur.375..659T,2003MNRAS.340L..28F,2007ARA&A..45..441M,2020arXiv201108948R}. Data over the last decade have revealed time delays between light curves observed at different energies that have been attributed to relativistic reverberation \citep{2009Natur.459..540F,2012MNRAS.422..129Z,2014A&ARv..22...72U,2016MNRAS.462..511K,2020NatAs...4..597A}, where spectral bands dominated by relativistic reflection are delayed relative to bands dominated by the continuum. Delays of bands dominated by the iron line at 6.4 keV (hereafter \FeK) in particular have provided some of the clearest evidence for these effects \citep{2012MNRAS.422..129Z,2016MNRAS.462..511K}. These reverberation signatures provide a potentially powerful probe of both accretion at the scale of the black hole horizon and gravity in the strong limit.

In recent years, we have been following up some of the highest signal reverberation measurements in the \FeK band with deeper exposures with many telescopes. These observations were optimized to confirm the early measurements and test simple predictions for the relativistic reverberation model.

In \cite{2019ApJ...884...26Z}, we presented new \xmm observations of the brightest Seyfert in the \FeK band, NGC 4151, which was the first object to show reverberation signal in the \FeK band. The energy-dependent lag at the shortest time scales in the new data does not resemble a simple broad iron line, unlike in the old data. Additionally, the new data showed no evidence for the relativistic reflection in the time-averaged spectra. These two results implied that the lags in NGC 4151 are unlikely to be due to reverberation of a relativistic component, but rather they could be due to variations in the complex absorbing system.

Reverberation delays in NGC 5506, another bright object, measured in archival \xmm data \citep{2016MNRAS.462..511K}, were also not confirmed in new data \citep{2020ApJ...893...97Z}. The low flux in the new data, however, meant that lags similar to those observed in the old data cannot be ruled out. By modeling both the spectral and timing signatures, it was found that the reflection fraction needed to explain the lags is larger than observed in the time-averaged spectrum, ruling out both a static lamp-post and simple wind reverberation models.

As a continuation of these tests, we present here new \nustar observations that test the predictions of the relativistic reverberation model in the Compton hump. The hump is a characteristic spectral feature of the reflection spectrum that peaks at 20--30 keV. The feature is produced in the spectrum by the competing effects of the photoelectric absorption above the iron edge at 7.1 keV, and the Compton scattering of the reflected photons, producing a local peak \citep{1991MNRAS.249..352G}. Under the relativistic reverberation hypothesis, light curves in the 20--30 keV band should be more delayed than those below $\sim20$ and above $\sim 40$ keV. \nustar data in two sources, \mcg and \swlong (hereafter \sw), suggested that this may be the case \citep{2014ApJ...789...56Z,2015MNRAS.446..737K}. Here, we report on the lag measurements from longer observations of the two sources taken in 2015 and 2018 respectively. We specifically focus in these tests on the \FeK and Compton hump lags because they are the {\it unambiguous} signatures of reflection, unlike the soft excess, whose origin is still highly debated \citep[e.g.][]{2012MNRAS.420.1848D,2013A&A...549A..73P,2019ApJ...871...88G}.

\begin{figure*}
    \centering
    \includegraphics[height=120pt]{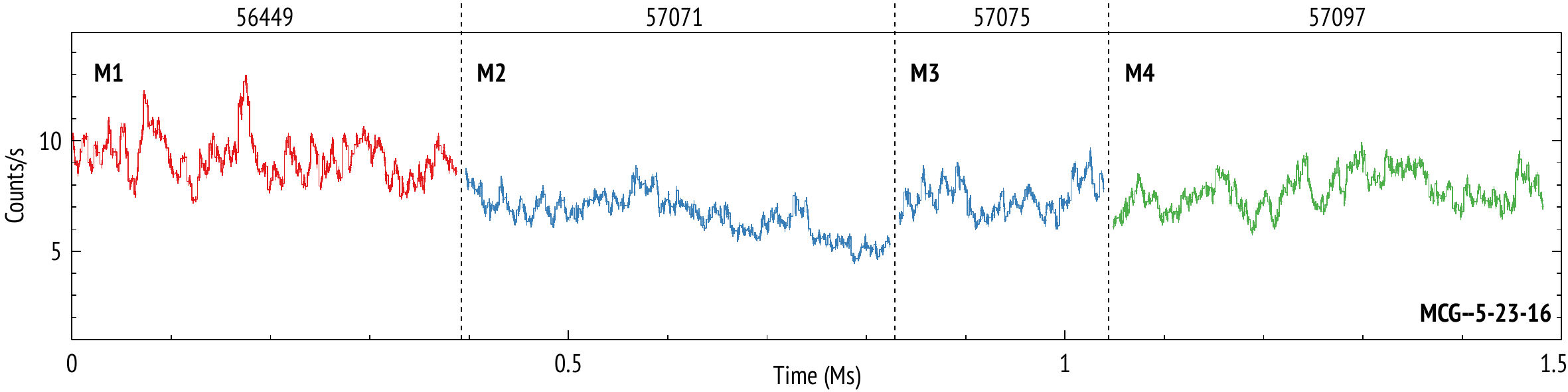}
    \includegraphics[height=120pt]{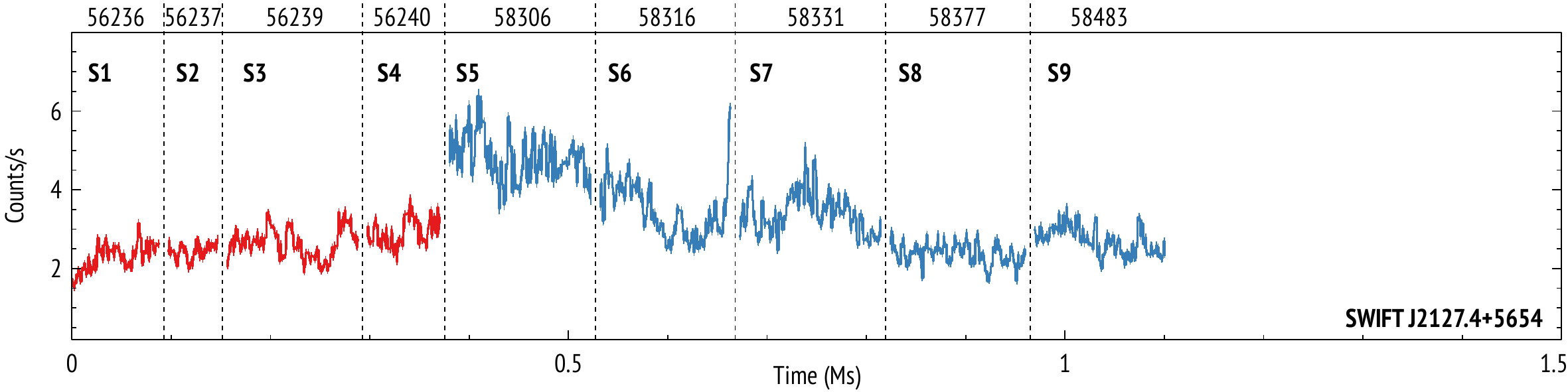}
    \caption{3--79 keV light curves from all observations considered in this study. The top panel is for \mcg and the bottom is for \sw. Individual observations are separated by the vertical dashed line. Different colors are used for the data groups used in section \ref{sec:results}. Note that not all light curves are contiguous and that the time plotted as a continuous variable for display clarity. The date in MJD units for the middle of each observation is shown in the top axis in each case.}
    \label{fig:lc}
\end{figure*}

\section{Observations \& Data Reduction} \label{sec:data}
\mcg was observed with long exposures 4 times by \nustar (obs-ids: 6000104600[2,4,6,8]; labeled M1--M4). The first was in 2013, and it was the first case where reverberation in the Compton hump was reported \citep[][hereafter \citetalias{2014ApJ...789...56Z}]{2014ApJ...789...56Z}. The spectral analysis of all the observations was reported in \cite[][hereafter \citetalias{2017ApJ...836....2Z}]{2017ApJ...836....2Z}.

\sw was observed in two programs, one in 2012 (obs-ids: 6000111000[2,3,5,7]; labeled S1--S4) and one in 2018 (obs-ids: 604020080[02,04,06,08,10]; labeled S5--S9). The spectral analysis of the first set was published in \cite{2014MNRAS.440.2347M}, and the timing analysis where the Compton hump reverberations were measured in \cite{2015MNRAS.446..737K}. Analysis of the data from the second set is presented here for the first time.

In both cases, the data were reduced and analyzed using the NuSTAR data analysis software, which is part of {\sc heasoft v6.27.2} . The data were reduced by running the standard pipeline {\tt nupipeline}. Source photons were then extracted for modules A and B from circular regions 2.5 arcmin in radius centered on
the source. Background photons were extracted from multiple regions around the source. The data were calibrated using {\sc caldb} release v20200626. For the timing analysis, light curves with bins of 512 seconds were extracted in 22 energy bins using {\tt nuproducts}. The energy bin boundaries are taken from two grids with logarithmic spacing, 14 bins between 3--10 keV and 7 between 10--79 keV. This results in the following bin boundaries in keV: 3, 3.3, 3.6, 4, 4.4, 4.8, 5.2, 5.7, 6.3, 6.9, 7.6, 8.3, 9.1, 10, 11.7, 13.8, 16.2, 19, 22, 31, 42, 58, 79. We do the lag calculations using both these bins (22 bins) and a coarse version where light curves in every two neighboring bins are summed first (producing 11 energy bins), before calculating the lags.

The lags are calculated using the method outlined in \cite{2013ApJ...777...24Z}, which was first presented in \cite{2010MNRAS.403..196M}, that uses piecewise models for the power spectra and transfer function to model the covariance of a light curve at some band and a reference light curve (taken to be the sum of all other bands minus the band of interest). The piecewise models are used to construct a covariance matrix and a Gaussian likelihood function, which is maximized numerically to obtain the lag values. 

In other words, the autocovariance is defined as:
\begin{equation}
\mathcal{A} (\tau) = {\displaystyle \int_{-\infty}^{+\infty} P(f) {\rm cos}(2\pi f \tau)\,df}
\end{equation}
where $f$ is the Fourier frequency, $\tau$ is the time difference between pairs of time points, and $P(f)$ is the power spectral density model, modeled here as a piecewise sum over $n_f$ frequency bins: $P(f) = \sum_i P_i$. Likewise, the cross-covariance is defined as:

\begin{equation}
\mathcal{X} (\tau) = {\displaystyle \int_{-\infty}^{+\infty} P(f)|\Psi_f| {\rm cos}(2\pi f \tau - \phi_f)\,df}
\end{equation}
where $|\Psi_f|$ and $\phi_f$ are the amplitude and phase of the transfer function that links the two light curves. These two are also modeled as piecewise functions of frequency ($\Psi_f = \sum_i \Psi_i$ and $\phi_f = \sum_i \phi_i$). Assuming the observed light curves are generated by such general Gaussian Process, a likelihood function can be constructed, and maximized to obtain best values $\Psi_i$ and $\phi_i$ as a function of frequencies $f_i$. The time lag in seconds is then obtained by dividing $\phi_i$ by $2\pi f_i$.

When calculating the power spectra and the time lag using some reported frequency bins, we also include two bands at the lowest and highest end of the observed frequency range. The lowest frequency bin extends to 0.5 times the lowest observed frequency (which is equal to $1/T$, where $T$ is the length of the longest light curve in seconds), and the highest frequency bin extends to 2 times the Nyquist frequency. Simulations show that these boundary bins are generally biased, and so their parameters are not reported.

The uncertainties in the reported parameters are obtained by running Monte Carlo Markov Chains starting with a random set of parameters around the best fit obtained by numerical maximization of the likelihood function. We use the $1\sigma$ uncertainty of the lag as the standard deviation of the parameter chain. The codes used for the analysis can be found in \url{https://github.com/zoghbi-a/fqlag}.

\begin{figure*}
    \centering
    \mcg\\
    \includegraphics[height=140pt]{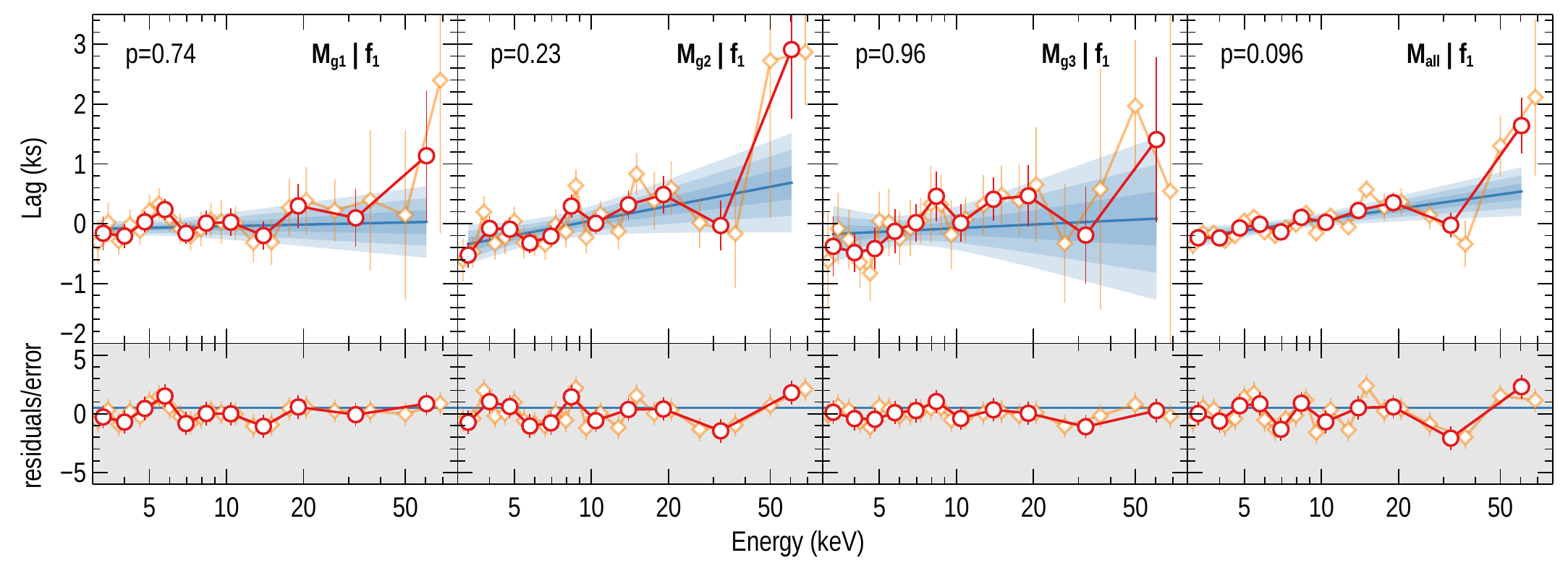}\\
    \sw\\
    \includegraphics[height=246pt]{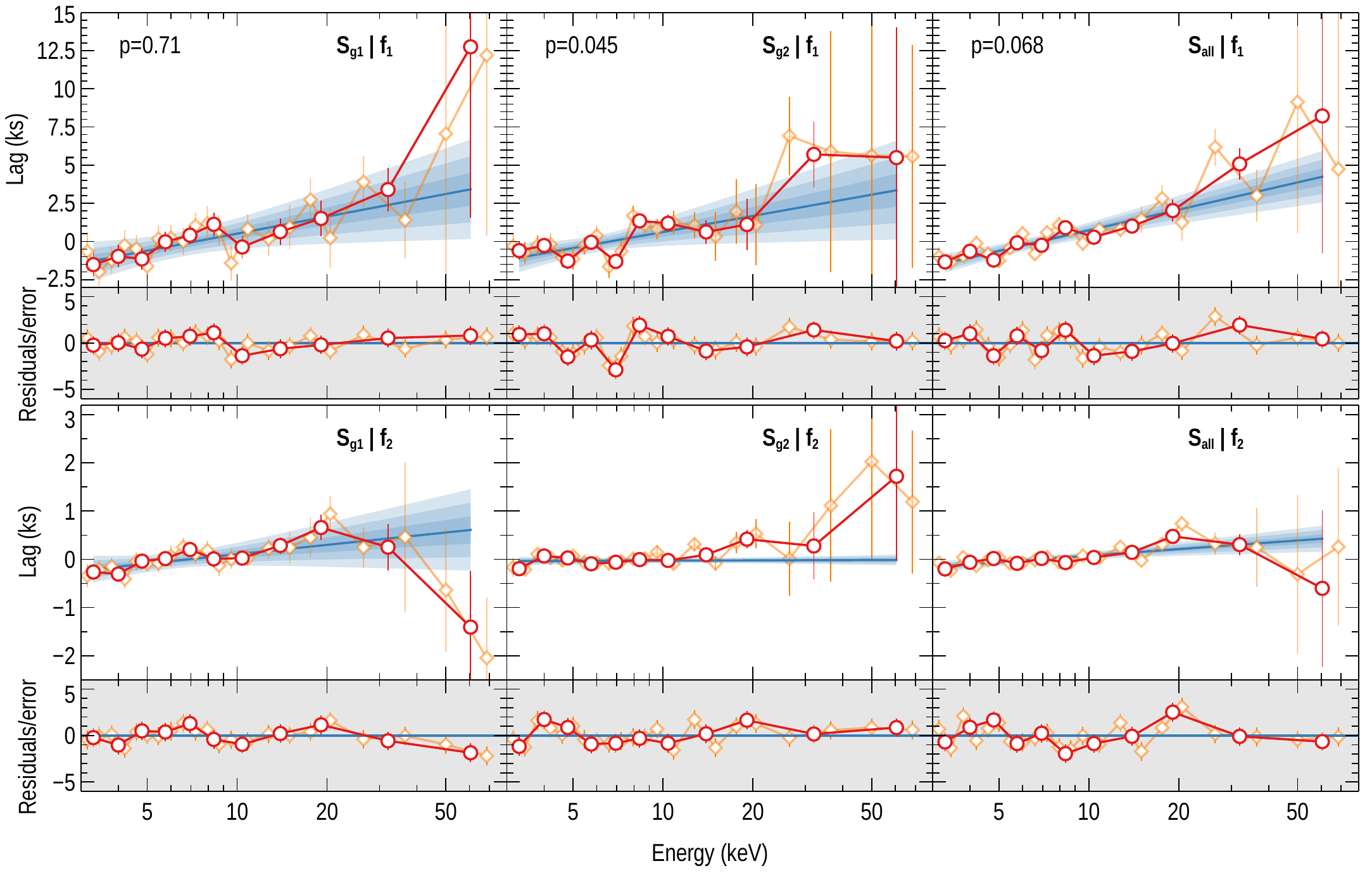}
    \caption{Energy-dependent time lags for \mcg (Top) and \sw (Bottom). In each panel, the lags from the coarse energy bins are shown as red circles, while the lags from the finer energy bins are shown in orange circles. The panel labels (shown in the top right of each panel) include M or S for the \mcg and \sw, respectively, the group number ($g1-g3$), and the frequency bin ($f_1$ or $f_2$). The numbers in the top-left corner are the null hypothesis probabilities when fitting log-linear models to the energy-dependent lags using the coarse bins (fitting the two frequencies simultaneously for the case of \sw). The best fit log-linear model is shown as a blue line, with the $1-3\sigma$ spread due to parameter uncertainties shown in blue shades. The residuals to this model from the coarse and fine energy bins are shown in the smaller panels with the grey background. The frequencies used for the case of \mcg are $(0.1-6)\times10^{-4}$ Hz. For \sw, the two frequency bins used are $(0.1-0.4)\times10^{-4}$ and $(0.4-4.5)\times10^{-4}$ Hz, and they are modeled simultaneously (with a single p-value)}. The parameters of the best fitting log-linear model are shown in Table \ref{tab:loglinear}.
    \label{fig:lag}
\end{figure*}

\section{Analysis \& Results}\label{sec:results}
Figure \ref{fig:lc} shows the 3--79 keV light curves from all the datasets used in the analysis. The combined light curve lengths exceed 1 Million seconds for each source. The figure shows that both sources exhibit strong variability on short time scales (sub 50 ks). Visual inspection of the light curves suggests that the characteristic time scale for \mcg is about 20--30 ks, while it is likely shorter ($\sim10$ ks) for \sw. We estimate the characteristic time scale by modeling the power spectra of the light curves with a bending powerlaw model \citep{2004MNRAS.348..783M}. This is a model that is found to describe the power spectra of AGN well, and it is often used to estimate the characteristic variability time scale from the frequency of the bend \citep[e.g.][]{2002MNRAS.332..231U}. On time scales shorter than the bend, the variability power drops rapidly and so it is natural to associate the bend frequency with some cut-off, or edge, in the accreting system, with indications that it is related to the viscous time scale in the inner accretion disk \citep{2004MNRAS.348..783M}. The piecewise spectral density values discussed in section \ref{sec:data} are constrained to follow a bending powerlaw model, whose parameters are again obtained by maximizing the likelihood function. We find bend frequencies (in Hz) of:
${\rm log}(f_b) = -4.44\pm0.10$ and $-4.09\pm0.19$ 
for \mcg and \sw, corresponding to time scales of around $27.5$ and $12.2$ ks, respectively.

\subsection{Delays at Published Frequencies}\label{sec:pub_freq}
The results of the inter-band time lags are shown in Figure \ref{fig:lag} for both \mcg (Top; panels labeled with M) and \sw (Bottom; labeled with S). These are delays of light curves of individual energy bands relative to a reference band. We take the reference light curve to be that in the total 3--79 keV band, subtracting the light curve of the band of interest each time. The lag is therefore zero at the average lag of the reference band (i.e. the whole band), weighted by count rate ($\sim$7 keV for both sources), and all other bands are relative to that energy.

For the case of \mcg, the lags are shown for one frequency bin, $(0.1-6)\times10^{-4}$ Hz, which is the bin in which a reverberation signal has been reported before \citepalias{2014ApJ...789...56Z}. The lag is shown for all the light curves together in the rightmost panel (M$_{\rm all} | {\rm f}_1$), and for 3 groups of observations, where ${\rm M}_{g1}$ and ${\rm M}_{g3}$ are individual observations ${\rm M}_1$ and ${\rm M}_4$ respectively, and ${\rm M}_{g2}$ combines ${\rm M}_2$ and ${\rm M}_3$ which are separated by less than a day. The groups are identified with different colors in Figure \ref{fig:lc}. The lags are plotted for both the coarse and full energy bins defined in section \ref{sec:data}. 

The basic model that we test the data against is a log-linear model of the form:
\begin{equation}\label{eqn:loglin}
    \tau = \alpha\times{\rm log}\left(\frac{E}{E_{\rm ref}}\right) \left(\frac{f}{10^{-4}}\right)^{-\beta}
\end{equation}
This takes into account both the energy and frequency dependence of the continuum lag, and it has been shown that it describes observed continuum lags in many sources \citep{2017MNRAS.468.3568E}. $\alpha$ and $\beta$ are model parameters, $f$ is the geometric center of the frequency bin, and $E_{\rm ref}$ is the reference energy band. For our case, the whole 3--79 keV band is used as a reference, so $E_{\rm ref}$ needs to be included as a free parameter. If one uses a single band as a reference, the model then has one less degree of freedom from the zero-lag at the reference band, and at the same time, $E_{\rm ref}$ is fixed, so the total number of degrees of freedom is the same.

This continuum model is motivated by observations of variability in stellar-mass black holes in Galactic X-ray binaries. Lags of the form shown in equation \ref{eqn:loglin} are observed to dominate the inter-band delays, and they have been shown to be associated with the primary spectral component that has a power-law shape in the time-averaged spectra. \citep{1989Natur.342..773M,1999ApJ...510..874N,2001MNRAS.327..799K}. In AGN, the same log-linear component is also observed at low frequencies \citep{2007MNRAS.382..985M,2017MNRAS.468.3568E}, and it was shown to be present in sources where the time-averaged spectra contain only a power-law component from the corona \citep{2013ApJ...777L..23W}. Several models have been discussed in the literature for the origin of the continuum lags, including Comptonization \citep{1989Natur.342..773M,1999ApJ...519..750K} and propagation models \citep{2001MNRAS.327..799K}.

Therefore, given that both the time-averaged spectra and the variability in the two sources under study here are dominated by the primary X-ray source from the corona, it is natural to use this log-linear model as a base for the modeling. In section \ref{sec:other_freq_rev_models}, we will also test the reverberation models with the unlikely possibility that the continuum lag component is not present in the data.

When modeling the energy-dependent lags from a single frequency bin, similar to the case of \mcg, the parameter $\beta$ cannot be constrained, so we fix it at $\beta=1$, as obtained from the modeling of other sources \citep{2017MNRAS.468.3568E}, though other values only shift the model normalization. The best-fit model for each data group is plotted in Figure \ref{fig:lag}, along with the 1, 2, and $3\sigma$ model uncertainty as shaded regions, obtained from the uncertainties in the best fit parameters, and the model parameters are summarized in Table \ref{tab:loglinear}.

\begin{table}[]
    \centering\footnotesize
    \begin{tabular}{|c|c|c|c|c|}
        \multicolumn{5}{c}{\mcg} \\ \hline\hline
                & ${\rm M}_{g1}$   & ${\rm M}_{g2}$   & ${\rm M}_{g3}$   & ${\rm M}_{all}$\\
         \hline
         $\alpha$ (sec)      &  $130\pm120$ & $570\pm170$ & $698\pm508$ & $387\pm87$\\
         $\beta$             & $1^\star$    & $1^\star$   & $1^\star$   & $1^\star$ \\
         $E_{\rm ref}$ (keV) &  $8.1\pm0.8$ & $8.1\pm0.9$ & $8.0\pm0.1$ & $7.8\pm0.9$\\
         $\chi^2 (9\; {\rm d.o.f}$)& 6.0  & 11.7        & 3.1         & 14.8 \\
         \hline
    \end{tabular}
    \begin{tabular}{|c|c|c|c|}
        \multicolumn{4}{c}{\sw} \\ \hline\hline
                & ${\rm S}_{g1}$   & ${\rm S}_{g2}$   & ${\rm S}_{all}$\\
         \hline
         $\alpha$ (sec)      &  $983\pm275$ & $368\pm183$ & $683\pm508$ \\
         $\beta$             & $0.9\pm0.2$  & $1.5\pm0.4$ & $1.2\pm0.1$ \\
         $E_{\rm ref}$ (keV) &  $6.9\pm0.7$ & $6.5\pm0.7$ & $6.8\pm0.4$ \\
         $\chi^2 (19\; {\rm d.o.f}$)& 15.2  & 30.6  & 28.9         \\
         \hline
    \end{tabular}
    
    \caption{Summary of the log-linear model of the form shown in equation \ref{eqn:loglin}. Plots of the data and the corresponding models are shown in Figure \ref{fig:lag}.}
    \label{tab:loglinear}
\end{table}


The p-values corresponding to the null hypothesis rejection probability are shown in the top left corner of each panel. The rejection of this continuum model would imply that the data requires other sources of delay, such as atomic features or additional components in the spectrum. Taking $p=0.05$ as a threshold for rejecting the model, the numbers in Figure \ref{fig:lag} for \mcg indicate that the data do not support any lag features beyond this log-linear continuum. 

A similar plot for \sw is also shown in Figure \ref{fig:lag} for two frequency bins, $(0.1-4)\times10^{-4}$ and $(4-4.5)\times10^{-4}$ Hz. These are the frequencies used in \cite{2015MNRAS.446..737K} analyzing data group ${\rm S}_{g1}$, where it was found that in the first frequency bin, the lags have a smooth log-linear shape, while for the second frequency bin, relativistic reverberation from both the iron line and Compton hump has been reported. The same model from equation \ref{eqn:loglin} is used to fit the data, but now because we have more than one frequency bin, we allow $\beta$ to be a free parameter, and the two frequency bins are modeled simultaneously.

The null hypothesis rejection probabilities shown in Figure \ref{fig:lag} indicate that the continuum log-linear model is sufficient in explaining the data of the first group (${\rm S}_{g1}$), but not the second group (${\rm S}_{g2}$). This suggests that additional model complexity beyond the log-linear continuum is needed in the latter case, and it is mostly driven by the complex residuals between 5--10 keV and the large lags around 40 keV shown in the residuals plots of panel ${\rm S}_{g2}|f_1$ in Figure \ref{fig:lag}. We note that these residuals are different from those seen in the \xmm data \cite{2015MNRAS.446..737K}.

\begin{figure}
\centering
    \mcg\\
    \includegraphics[height=120pt]{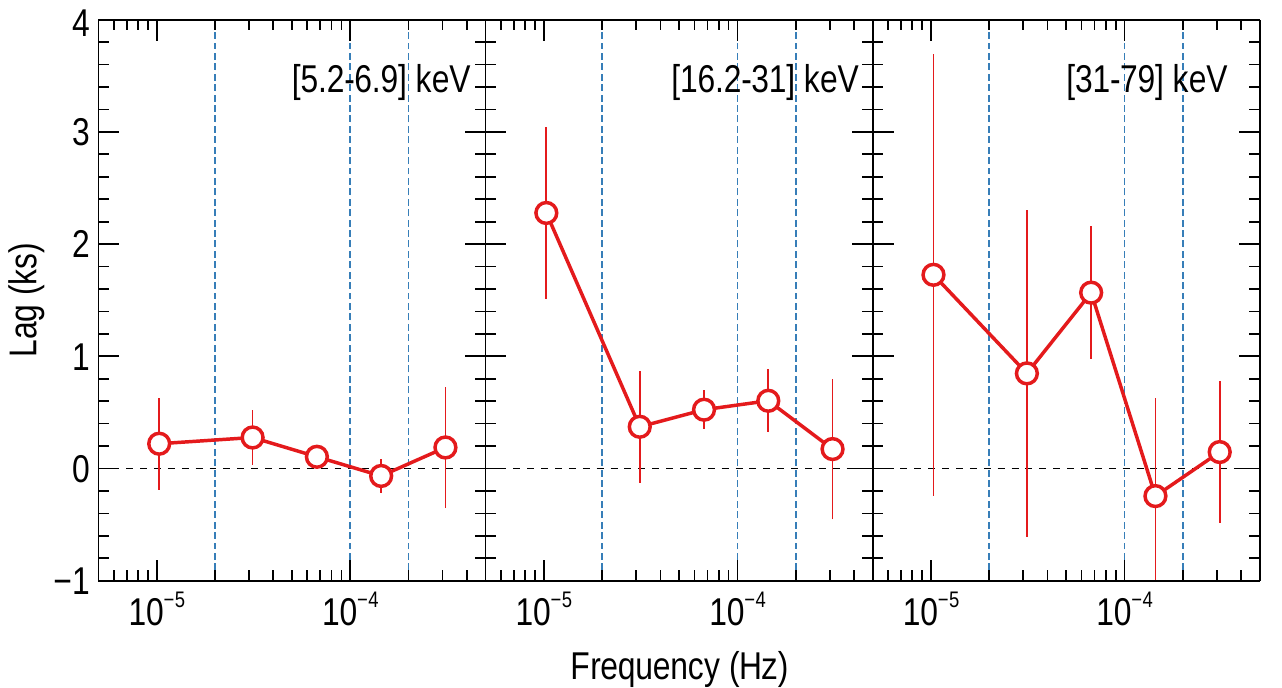}\\
    \sw\\
    \includegraphics[height=120pt]{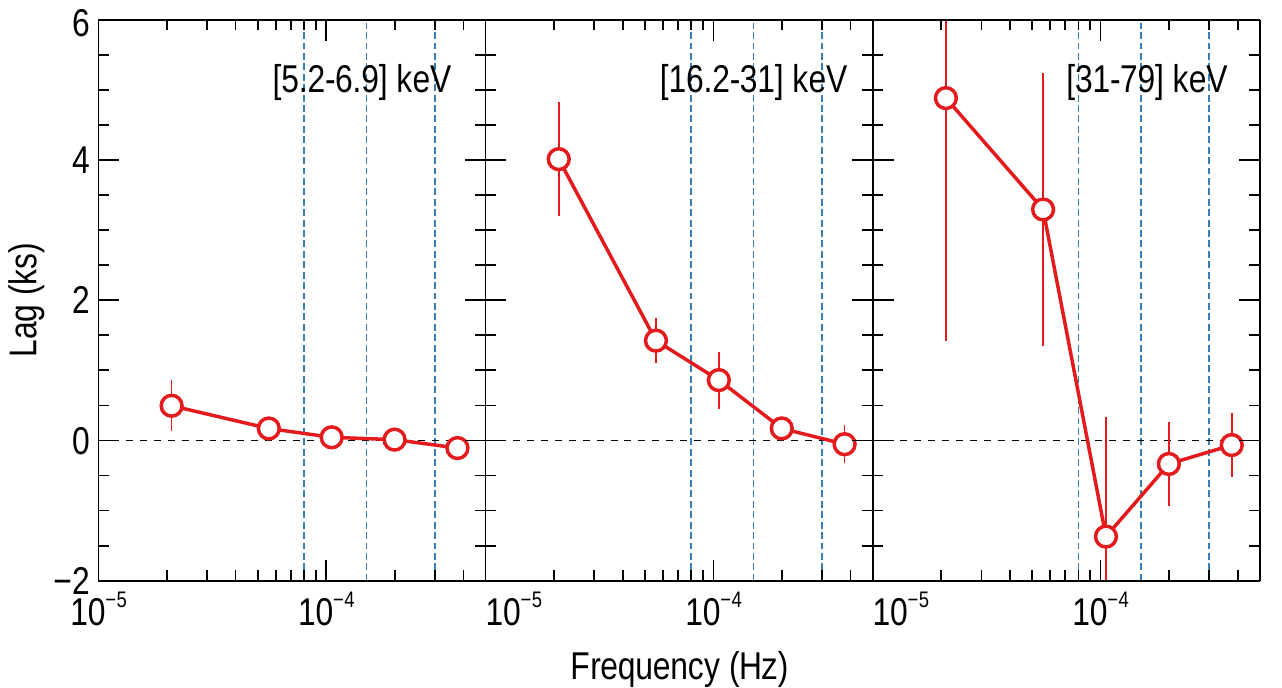}
    \caption{Frequency-dependent time lags for \mcg (Top) and \sw (Bottom). The lags are shown for three energy bands: 5.2--6.9, 16.2-31, and 31--79 keV, which are shown in the three panels from left to right respectively. The lags are measured relative to the 3--5.2 keV band. The dashed vertical lines show the bin boundaries used in calculating the energy-dependent lag.}
    \label{fig:lag_fq}
\end{figure}

\subsection{Delays at Other Frequencies}\label{sec:other_freq}
The results of the previous section show that the new data do not show lags at the {\emph same} frequency as previously observed, with the modeling implying that the observed lags can be explained by a continuum log-linear model, with no need for additional reverberation components. In this section, we do not restrict the analysis to the published frequencies, but rather, we examine all observed frequencies, and focus on data from all observations together, which provides the highest signal.

Figure \ref{fig:lag_fq} shows the lag as a function of Fourier frequency for three energy bands: 5.2--6.9, 16.2-31, and 31--79 keV, measured relative to the 3--5.2 keV band. The reference band is dominated by the continuum power law spectral model in the time-averaged spectra of both sources. The other bands are selected so they probe the iron line, the peak of the Compton hump, and the energy where the contribution of the Compton hump starts to diminish.

The figure shows that for the first two bands, there is a clear delay relative to the reference band (i.e. the bands at 5.2--6.9 and 16.2-31 keV lag the reference 3--5.2 keV band). This however cannot be associated with the iron line or the Compton hump because it is similar to what a continuum lag model predicts, where hard bands are predicted to lag softer bands. The answer may come from the energy-dependent lag, which we will explore next. The third band (31--79 keV) is potentially interesting. Here, the continuum lag model predicts that the lag should increase compared to the second band because the contribution of the power law component in the spectrum increases. The figure shows that this may be the case at frequencies less than $10^{-4}$ Hz. Above that, the lag appears to drop, suggesting that lags due to reverberation from the Compton hump may be contributing around $10^{-4}$ Hz. Although this is only suggestive, we explore it further with the energy-dependent lag discussed next.

\begin{figure*}
    \centering
    \begin{tabular}{c|c}
         \mcg &  \sw \\
         \includegraphics[width=250pt]{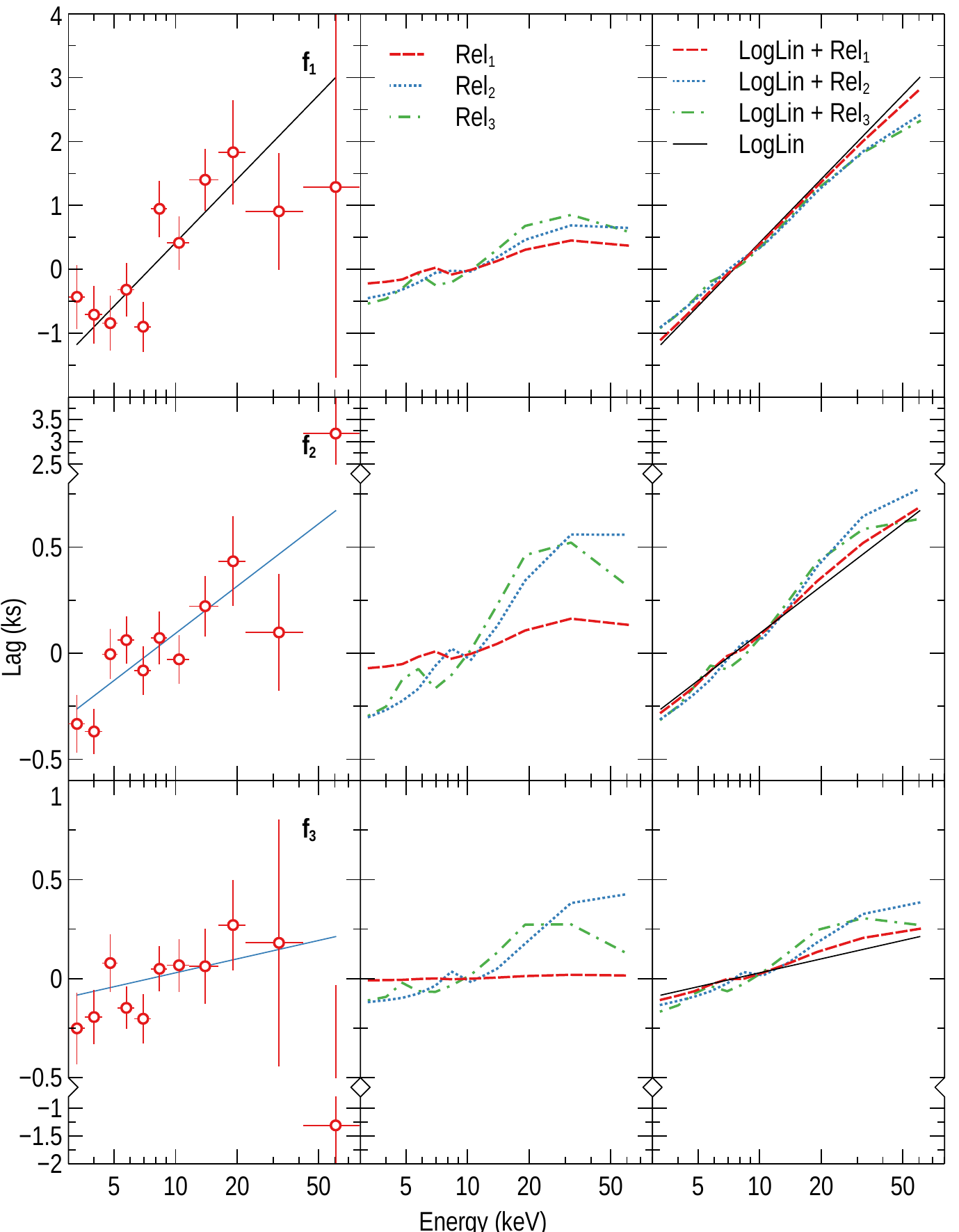} &
         \includegraphics[width=250pt]{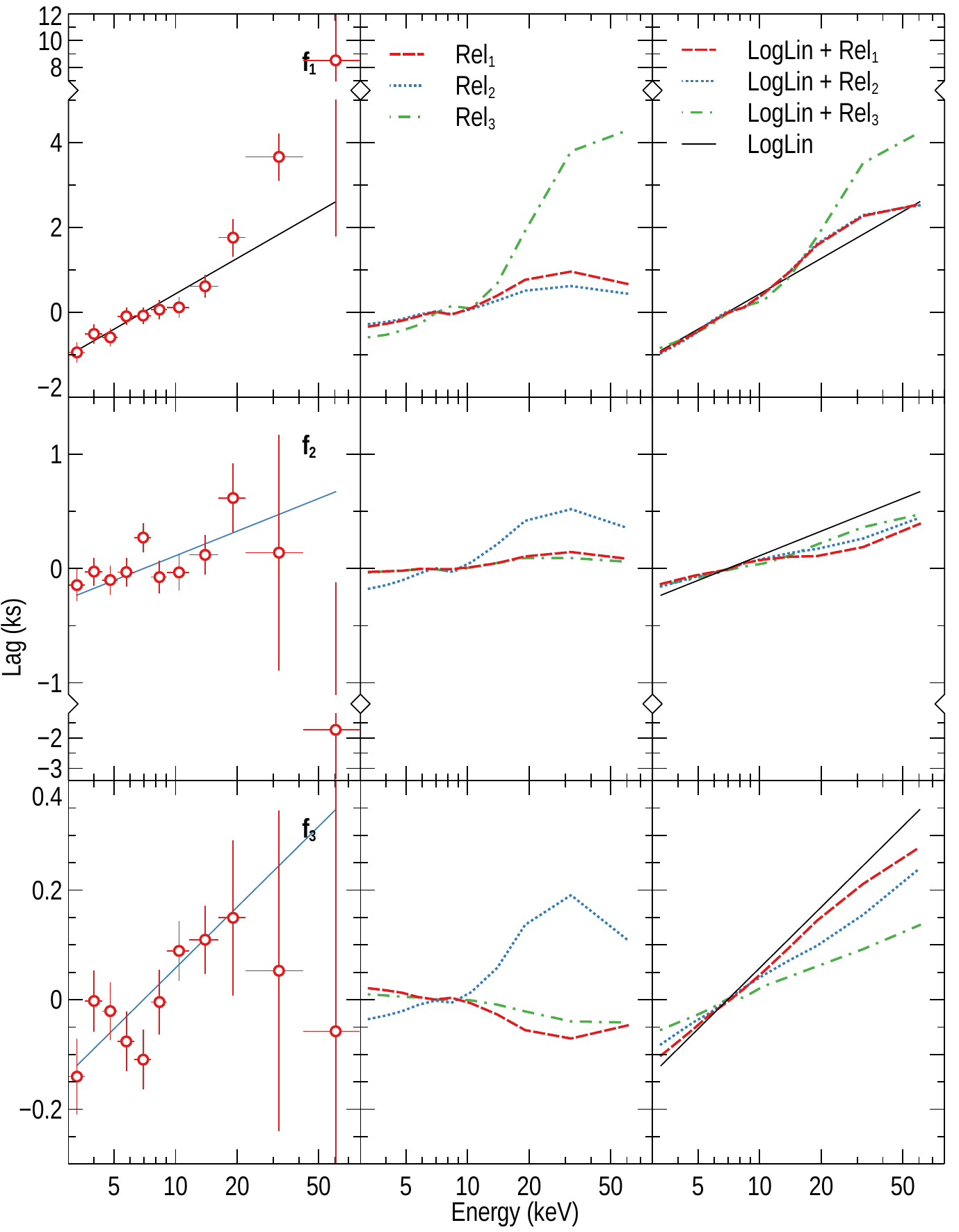} \\
    \end{tabular}
    \caption{Energy-dependent time lags for \mcg (Left) and \sw (Right). Each row represents the energy-dependent lag at one frequency bin. The frequency boundaries are $5\times10^{-6}, 2\times10^{-5}, 10^{-4}, 2\times10^{-4}$ Hz for \mcg, and $10^{-5}, 8\times10^{-5}, 1.5\times10^{-4}, 3\times10^{-4}$ Hz for \sw. The left column in each case shows the data along with the best fit log-linear model in solid blue color ({\tt LogLin}). The middle and right panels contain the best fit models discussed in the text plotted separately from the data for clarity. See the caption of Table \ref{tab:fit_e1} for model definitions. Note the broken axes, which are used to ensure features in both the data and models are visible when plotted on the same scale. The three frequencies in each case are modeled simultaneously as discussed in section \ref{sec:other_freq}.}
    \label{fig:lag_e1}
\end{figure*}

Based on the frequency dependence, we calculate the energy-dependent lags at three frequency bins for each source, defined by the following boundaries: $5\times10^{-6}, 2\times10^{-5}, 10^{-4}, 2\times10^{-4}$ Hz for \mcg, and $10^{-5}, 8\times10^{-5}, 1.5\times10^{-4}, 3\times10^{-4}$ Hz for \sw. The frequency bin boundaries are plotted as dashed vertical lines in Figure \ref{fig:lag_fq}. The resulting energy-dependent lags are shown in Figure \ref{fig:lag_e1}.

The data plots in Figure \ref{fig:lag_e1} show features that suggest there may be deviations from a simple log-linear continuum model. To model this lag spectra, we use a combination of a log-linear model for the continuum and a relativistic reverberation model incorporating different assumptions. For the latter, we specifically test the data against the predictions of the lamp-post relativistic model, where we use {\tt kynxilrev}\footnote{Available from \url{https://projects.asu.cas.cz/stronggravity/kynreverb}}, which computes the time-dependent reflection spectra of the disc as a response to a flash of primary power-law radiation from a point source located on the axis of the black-hole accretion disc \citep{2004ApJS..153..205D}. This is similar to other models that have been used before in other work \citep{1995MNRAS.272..585C,1999ApJ...514..164R,2013MNRAS.430..247W,2014MNRAS.439.3931E,2014MNRAS.438.2980C,2015MNRAS.452..333C,2019MNRAS.488..324I}.

In all the following model fits, the lag-energy data are modeled across all three frequencies simultaneously. For the log-linear model, this is achieved by using the frequency-dependence form in equation \ref{eqn:loglin}, while {\tt kynxilrev} includes the frequency dependence self-consistently as inferred from the calculations of the relativistic response function. The model parameters and fit statistics are summarized in Table \ref{tab:fit_e1}.

The data are first tested against the simplest model, a log-linear function (equation \ref{eqn:loglin}) similar to that used in section \ref{sec:pub_freq}. This is referred to as {\tt LogLin} in Figure \ref{fig:lag_e1} and Table \ref{tab:fit_e1}. This model accounts for most of the variance in the data and provides an acceptable fit for both sources ($\chi^2/$d.o.f = 41/31 and 39/31 for \mcg and \sw, respectively). Note that the $\beta$ parameter is fixed at 1, the value obtained by modeling of \xmm data in other sources \cite{2017MNRAS.468.3568E}. Allowing it to vary provides only a small and insignificant improvement in the fit, and its value is consistent with 1, so in all subsequent analyses, we keep its value fixed at 1.

\subsection{Testing the Lamp-post model}

Next, we explore lamp-post relativistic reverberation models. Given the lag data quality and the complexity of these models, it is generally not possible for the lag data alone to constrain all the parameters that can affect the model \citep[e.g.][]{2014MNRAS.438.2980C,2015MNRAS.452..333C}. Additionally, any model that explains the reverberation lags needs to simultaneously be consistent with the time-averaged spectrum. Because data in the time-averaged spectra has a much high signal compared to the lag, different groups approach the simultaneous fitting differently \citep{2013MNRAS.430..247W,2015MNRAS.452..333C,2020MNRAS.498.4971M}. Here, we assume that the variable spectrum has a similar shape to the time-averaged spectrum (excluding the distant reflector). This assumption is specifically motivated by the fact that our goal is to check if the parameters inferred from the lag data are consistent with the time-averaged model, often used to infer the parameters of the lamp-post. A more precise procedure would be to model the lag and the cross-spectrum simultaneously \citep[e.g.][]{2020MNRAS.498.4971M}, but the uncertainties on the inferred parameters are generally larger than those inferred from the time-averaged spectrum.


\begin{table*}
\hspace{-1.7cm}
\footnotesize
    \begin{tabular}{|c|c|c|c|c|c|c|c|}
        \multicolumn{8}{c}{\mcg} \\ \hline\hline
        Spectral Model & \multicolumn{7}{c}{$a=0.998^\star$, $R_{\rm out} = 400^\star$, 
                        $\Gamma=1.79$, log$\xi=0.5$, $A_{\rm Fe}=1.14$, $E_c=137$} \\ \hline
                                &   {\tt LogLin}      & 
                                    {\tt Rel$_1$}    & {\tt LogLin + Rel$_1$}  & 
                                    {\tt Rel$_2$}    & {\tt LogLin + Rel$_2$} &
                                    {\tt Rel$_3$}    & {\tt LogLin + Rel$_3$} \\
         \hline
         $\alpha$ (sec)         & $331\pm50$    & -                 & $300\pm52$    &
                                 -              & $214\pm74$    & -                     &
                                 $216\pm82$     \\
         $\beta$                & $1^\star$     & -                 & $1^\star$     &
                                 -              & $1^\star$     & -                     &
                                 $1^\star$      \\
         $E_{\rm ref}$ (keV)    & $7.5\pm0.6$   & -                 & $7.4\pm0.7$   &
                                 -              & $6.8\pm1.0$   & -                     &
                                 $6.8\pm1.0$    \\
         $M (10^{7} M_{\odot})$ & -             & $0.5\pm0.3$       & $<0.3$        &
                                 $3^\star$      & $3^\star$     & $3.7\pm1.3$           &
                                 $1.3^{+2.3}_{-1.1}$    \\
         $R_{\rm in} $($r_g$)   &-              & $49^s$            & $49^s$        &
                                $3.3\pm0.9$     & $3.8^{+3.5}_{-0.8}$ & $4.7\pm0.9$  &
                                $4.9^{+1.8}_{-3.4}$     \\
         $h$ ($r_g$)            &-              & $86^s$            & $86^s$        &
                                $4.6\pm2.8$     & $<14$         & $2.4\pm0.9$       &
                                $2.4^{+6.3}_{-1.1}$          \\
         $\theta$ ($^\circ$)    &-              & $80^s$            & $80^s$        &
                                $80^s$          & $80^s$        & $<9$                 &
                                $<44$ \\
         \hline
         $\chi^2$ / d.o.f       &{\bf 41/31}    & 71/32             & 41/30         &
                                52/31           & 40/29         & 47/29                 &
                                38/27\\
         $p$-value              &{\bf 0.11}     & $10^{-4}$         & 0.092         &
                                $0.011$         & 0.08          & 0.018                 &
                                0.08\\
         \hline\hline
         
         \multicolumn{8}{c}{\sw} \\ \hline\hline
         Spectral Model & \multicolumn{7}{c}{$a=0.998^\star$, $R_{\rm out} = 400^\star$, 
                        $\Gamma=2.0$, log$\xi=1.7$, $A_{\rm Fe}=0.91$, $E_c=64$} \\ \hline
         
         $\alpha$ (sec)         & $784\pm86$    & -                 & $598\pm105$    &
                                 -              & $577\pm100$       & -                 &
                                 $365\pm62$     \\
         $\beta$                & $1^\star$     & -                 & $1^\star$     &
                                 -              & $1^\star$         & -                 &
                                 $1^\star$      \\
         $E_{\rm ref}$ (keV)    & $7.0\pm0.4$   & -                 & $7.1\pm0.5$   &
                                 -              & $7.0\pm0.5$       & -                 &
                                 $7.4\pm0.9$    \\
         $M (10^{7} M_{\odot})$ & -             & $5.5\pm1.0$       & $10^{+1}_{-4}$    &
                                 $1.5^\star$    & $1.5^\star$       & $17.1\pm0.3$       &
                                 $19^{+1}_{-5}$  \\
         $R_{\rm in} $($r_g$)   &-              & $4.7^s$           & $4.7^s$       &
                                $<3$            & $4.1^{+21}_{2.8}$ & $<1.9$ &
                                $<2.2$          \\
         $h$ ($r_g$)            &-              & $4.5^s$           & $4.5^s$       &
                                $9.6\pm1.3$    & $28^{+22}_{-9}$   & $2.4\pm0.2$       &
                                $2.4^{+0.5}_{-0.2}$     \\
         $\theta$ ($^\circ$)    &-              & $43^s$            & $43^s$        &
                                $43^s$          & $43^s$            & $70\pm4$    &
                                $68^{+12}_{-6}$  \\
         \hline
         $\chi^2$ / d.o.f       &39/31          & 75/32             & 31/30         &
                                73/31           & 31/29             & 32/29             &
                                {\bf 23/27}\\
         $p$-value              &0.16           & $3\times10^{-5}$  & 0.40          &
                                $3\times10^{-5}$& 0.37              & 0.31              &
                                {\bf 0.68}\\
        \hline\hline
         
    \end{tabular}
    
    \caption{Model fit parameters for the frequency and energy-dependent lags for \mcg (Top) and \sw (Bottom). The models are fit simultaneously to the three frequency bins plotted in Figure \ref{fig:lag_e1}. The model details are discussed in section \ref{sec:other_freq}. {\tt LogLin} is a log-linear continuum model of the form shown in equation \ref{eqn:loglin}. {\tt Rel$_1$} is a relativistic reverberation model where all the parameters are fixed to those in the time-averaged spectrum, and the black hole mass $M$ is allowed to vary. In {\tt Rel$_2$}, $M$ is fixed to literature values and the inner radius of the disk $R_{\rm in}$ and the height of the lamp-post $h$ are free parameters. {\tt Rel$_3$} is the most flexible of all the relativistic reverberation models, where $M$, $R_{\rm in}$, $h$ and the inclination of the inner disk $\theta$ are free parameters.\\
    $^\star$: The parameters is fixed.
    $^s$: The parameter is fixed to the time-averaged spectral model.
    For the spectral model parameters: $a$ is the black hole spin, $R_{\rm out}$ is the outer radius of the disk in $r_g$, $\Gamma$ is the photon index, $\xi$ is the ionization parameter, $A_{\rm Fe}$ is the iron abundance and $E_c$ is the cutoff energy in keV.
    }
    \label{tab:fit_e1}
\end{table*}

\subsubsection{The Time-Averaged Spectra}\label{sec:time-avg-fit}
The time-averaged spectra are extracted and modeled to obtain a spectral model. For \mcg, the spectral modeling of the data was presented in \citetalias{2017ApJ...836....2Z}. That modeling, however, allowed the disk emissivity and reflection fraction to be different from a lamp-post geometry, and the lag predictions with those assumptions are not unique. We, therefore, refit the spectra making the lamp-post geometry assumption explicitly.

We therefore model the \nustar spectra of both \mcg and \sw with {\tt relxilllp} \citep{2014ApJ...782...76G,2014MNRAS.444L.100D}. Spectra of the two sources from \xmm PN were also included in the analysis. The higher energy resolution of the PN helps in separating the broad and narrow iron line components, and also in measuring the iron absorption edge at 7.1 keV. Archival PN data are reduced following the standard procedures \citepalias[e.g.][]{2017ApJ...836....2Z}, and we use observation IDs (0302850201, 0727960101, 0727960201) for \mcg and (0601741901, 0655450101, 0693781701, 0693781801, 0693781901) for \sw. The total spectral model includes also (in addition to {\tt relxilllp}) {\tt xillver} to model the distant reflection, and {\tt zTBabs} to model neutral absorption in the host galaxy. 

Because the data were not simultaneous, we initially assumed that only the photon index and the intensity of the primary power law continuum change between observations, and then we explored allowing other parameters to vary. We found that for both sources, allowing the intensity of the distant reflection model to vary provides the most significant fit improvement. The final fit statistics $\chi^2/{\rm d.o.f}$ are: 1593/1468 and 2056/1907 for \mcg and \sw respectively. The parameters that are relevant for the lag modeling are summarized in the top row for each source in Table \ref{tab:fit_e1}.

\subsubsection{Relativistic Reverberation Models}\label{sec:other_freq_rev_models}
The first relativistic model we explore is one where all the geometry parameters are assumed to be similar to the time-averaged spectra, with the black hole mass being the only variable parameter, which sets the physical scale for the model. This is model {\tt Rel$_1$}. In {\tt Rel$_2$}, we instead fix $M$ at known estimates for the mass ($3\times10^7$ and $1.5\times10^{7} M_\odot$ for \mcg and \sw, respectively), and allow the inner radius of the disk $R_{\rm in}$ and height of the corona $h$ to vary. For both of these cases, the model is significantly rejected by the data for the two sources (see the p-values in Table \ref{tab:fit_e1}). The two models are plotted in the middle column of each panel in Figure \ref{fig:lag_e1}. This result implies that a reverberation model {\it alone} cannot explain the lag data in both sources, even when allowing for $R_{\rm in}$ and $h$ to be different from those of the time-averaged spectrum.

In model {\tt Rel${_3}$},  we allow $R_{\rm in}$, $h$ and $M$ to vary at the same time, and we additionally allow the inclination $\theta$ to vary too. We find that this model can be also be significantly rejected for the case of \mcg ($p=0.018$), but it is consistent with the data for \sw ($p=0.31$). The best-fitting parameters in the latter deviate significantly from those obtained in the time-averaged spectrum. This inconsistency is discussed further in section \ref{sec:consist}.

\subsubsection{Continuum \& Relativistic Reverberation Models}
For each of the three relativistic models, we next add a log-linear continuum, and summarize the parameters in Table \ref{tab:fit_e1} (columns {\tt LogLin + Rel${_1}$}, {\tt LogLin + Rel${_2}$} and {\tt LogLin + Rel${_3}$}). The total model in each case is the sum of the two models because the convolution of the two response functions in the time domain corresponds to the sum of the phases \citep[e.g.][]{2014A&ARv..22...72U}.

For \mcg, the log-linear model on its own already provides a good description to the data, and none of the additional relativistic reverberation models provide any significant improvement to the fit. Because the log-linear model with and without the relativistic reverberation models are nested, we use the F-test to assess the improvement significance of the former compared to the latter. The cases of {\tt Rel$_1$} and {\tt Rel$_2$} provide no improvement in the fit, because the fit does not improve. For {\tt Rel$_3$}, the null hypothesis probability that the addition of the new model provides {\it no} improvement is 0.71, i.e. the additional {\tt Rel$_3$} model provides no significant improvement compared to a log-linear continuum.
Plots of theses models in the third column of Figure \ref{fig:lag_e1} show only small deviation from the continuum-only model ({\tt LogLin}).

\begin{table}[]
    \centering\footnotesize
    \begin{tabular}{|c|c|c|c|c|}
        \multicolumn{5}{c}{\mcg} \\ \hline\hline
                & ${\rm M}_{g1}$   & ${\rm M}_{g2}$   & ${\rm M}_{g3}$   & ${\rm M}_{all}$\\
         \hline
         $\alpha$ (sec)         &  $290\pm117$ & $407\pm110$ & $334\pm180$ & $331\pm50$\\
         $\beta$                & $1^\star$    & $1^\star$   & $1^\star$   & $1^\star$ \\
         $E_{\rm ref}$ (keV)    &  $7.3\pm1.7$ & $8.1\pm1.3$ & $7.5\pm2.6$ & $7.5\pm0.6$\\
         $\chi^2 (31\; {\rm d.o.f}$)& 20  & 24      & 18          & 41 \\
         \hline
    \end{tabular}
    \begin{tabular}{|c|c|c|c|}
        \multicolumn{4}{c}{\sw} \\ \hline\hline
                & ${\rm S}_{g1}$   & ${\rm S}_{g2}$   & ${\rm S}_{all}$\\
         \hline
         $\alpha$ (sec)      & $903\pm170$ & $584\pm181$ & $784\pm86$ \\
         $\beta$             & $1^\star$   & $1^\star$   & $1^\star$ \\
         $E_{\rm ref}$ (keV) & $7.2\pm0.7$ & $6.7\pm0.8$ & $7.0\pm0.4$ \\
         $\chi^2 (31\; {\rm d.o.f}$)& 22  & 44     & 39            \\
         \hline
    \end{tabular}
    
    \caption{Summary of the log-linear model for individual groups of observations the form shown in equation \ref{eqn:loglin} using three frequency bins defined in section \ref{sec:other_freq}.}
    \label{tab:loglinear_e1}
\end{table}

In \sw, the relativistic reverberation models provide a significant improvement in the fit in all three cases when compared to the {\tt LogLin} case ($\Delta\chi^2/{\rm d.o.f}$ of 8/1 for {\tt Rel$_1$} and {\tt Rel$_2$}, and 16/4 for {\tt Rel$_3$}). Although the continuum-only model gives a satisfactory description of the data, the fact that {\tt Rel$_{3}$} alone also describes the data well implies that neither of the two models ({\tt LogLin} and {\tt Rel$_3$}) can be rejected on its own. Their combination provides a significant fit improvement compared to either of them individually ($\Delta\chi^2/{\rm d.o.f}$ of 16/4 relative to {\tt LogLin} and 9/2 relative to {\tt Rel$_3$}). We therefore conclude that based on the fit statistics alone, the combination of the two models ({\tt LogLin} + {\tt Rel$_3$}) gives the best description of the \sw lag data.  

\subsection{Consistency Between The Time-Averaged \& Lag Models}\label{sec:consist}
The data of \mcg do not show any strong evidence for a relativistic reverberation component, so we focus in this section on \sw, whose best fit model includes a relativistic reverberation model, in addition to a log-linear continuum model. We want to examine in this section if the parameters from such model are consistent with the time-averaged spectra.

\begin{figure}
    \centering
    \includegraphics[height=190pt]{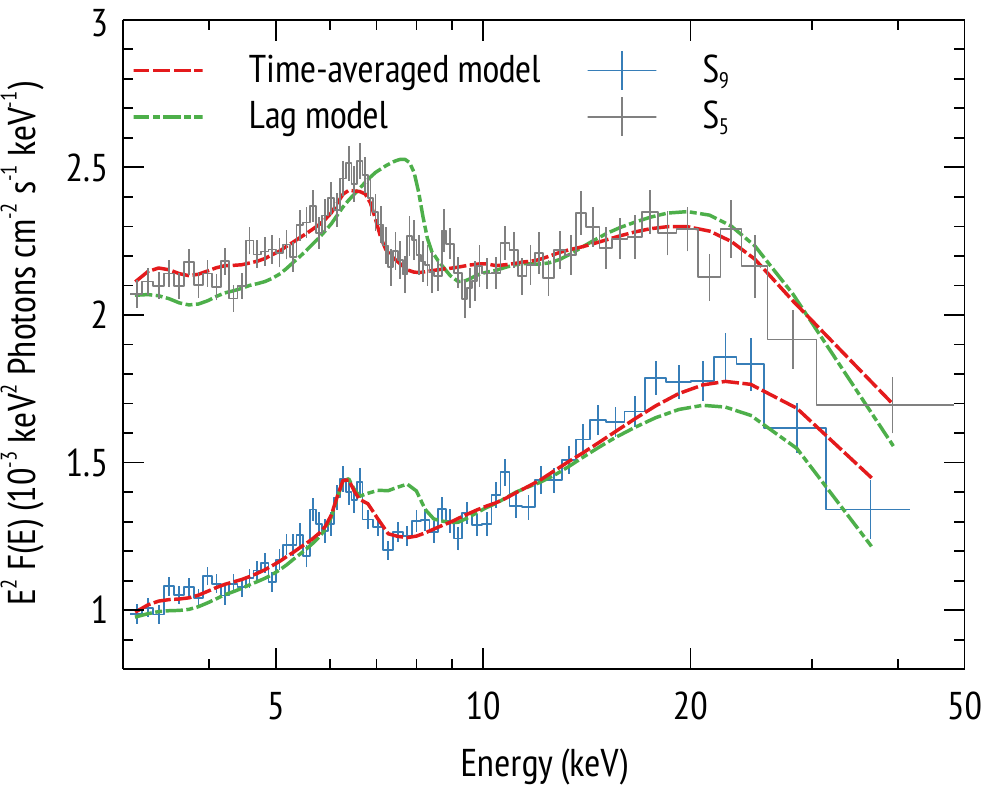}
    \caption{Time-averaged spectral models for \sw showing the difference between the the best fitting model to the time-averaged spectra (red dashed) and the model where $R_{\rm in}$, $h$ and $\theta$ are fixed to the best fit values from the lag data (green dash-dotted). The labels S$_5$ and S$_9$ refer to two separate observations as defined in Figure \ref{fig:lc} and section \ref{sec:data}.
    }
    \label{fig:consist}
\end{figure}

In Figure \ref{fig:consist}, we plot the time-averaged spectra from two observations {\tt S$_5$} and {\tt S$_9$} that sample different flux intervals, along with two models in each case (see Figure \ref{fig:lc} and section \ref{sec:data} for the label definitions). The first model fits the time-averaged spectra (from section \ref{sec:time-avg-fit}), and the second is the model predicted given the best fit relativistic reverberation model {\tt Rel$_3$}. As Table \ref{tab:fit_e1} shows, the best parameters for {\tt Rel$_3$} that fit the lag data alone or along with the log-linear model are almost identical, so we only plot the model from the latter. The plotted model in each case includes the contribution from both the relativistic reflection component and the primary power-law under the lamp-post assumption, along with {\tt xillver} that fits the narrow iron line. 

The second model is obtained by fixing $R_{\rm in}$, $h$ and $\theta$ to the values from the {\tt LogLin} + {\tt Rel$_3$} column in Table \ref{tab:fit_e1}, and then re-fitting all 14 (9 \nustar and 5 \xmm) time-averaged spectra simultaneously.

The high inclination required in the lag model produces stronger and more blue-shifted reflection features than those in the lower inclination case. In other words, the large inclination required to model the lag data predicts the iron line in the time-averaged spectra to peak closer to 8 keV, which differs significantly from the observation.

Additionally, the mass inferred in the lag model of $M=(1.9^{+0.1}_{-0.5})\times10^{8} M_\odot$ is an order of magnitude larger than the value of $1.5\times10^{7} M_\odot$, estimated from single epoch optical spectroscopy \citep{2008MNRAS.389.1360M}. The single-epoch mass estimate may have an uncertaintiy of a factor of a few, but it remains below the best fit value from the lag data. 

The mass can also be estimated using the excess variance \citep{2012A&A...542A..83P}. From the light curves, we calculate the excess variance in the 3--10 keV band for 80 ks segments and find a value of $\sigma^2_{\rm rms} = 0.024\pm0.001$, which gives a mass estimate of $(6\pm3)\times10^{6} M_{\odot}$, which is even smaller than the estimate from optical spectroscopy, and consistent with the classification of \sw as a narrow line Seyfert 1 nucleus. This estimate is however also not consistent with the value inferred from the lamp-post modeling of the lag. Note that the upper uncertainty of $+0.1$ is because we enforced an upper limit on the mass of $2\times10^{8} M_{\odot}$, as it is unlikely that the mass of \sw exceeds this limit given its classification as an NLS1.

\section{Discussion \& Conclusions}

\subsection{Summary \& Context}
Detecting and interpreting inter-band delays in AGN is crucial as it could provide a powerful probe of the 
accretion activity down at the black hole scale \citep{2009Natur.459..540F, 2012MNRAS.422..129Z, 2013MNRAS.430..247W,2014MNRAS.439.3931E,2014MNRAS.438.2980C,2016MNRAS.462..511K,2020NatAs...4..597A}. Focusing specifically on testing the prediction of the relativistic reverberation picture, we have targeted several of the brightest Seyfert Galaxies in X-ray with different observing programs to perform specific tests of the picture \citep{2019ApJ...884...26Z,2020ApJ...893...97Z}. Here, we are reporting the results from one additional study that focuses on the Compton Hump and hard X-ray band.

Relativistic reverberation signal that has been attributed to the Compton hump was reported first in \mcg \citepalias{2014ApJ...789...56Z} and later in \sw and NGC 1365 \citep{2015MNRAS.446..737K}. Our data almost triples the amount of analyzed exposure on the first two sources. Our main result is that reverberation delays are not confirmed in \mcg, and the data is consistent with a continuum model that has a log-linear form. The data rule out delays from a lamp-post model with parameters that produce the time-averaged spectrum.

In \sw, both a log-linear model and relativistic reverberation model can fit the lag data separately, and their combination provides a significant improvement in the fit compared to either of the models individually (Table \ref{tab:fit_e1}). The parameters of a lamp-post relativistic reverberation model are however not consistent with the time-averaged spectrum. The lamp-post model that describes the lag data implies a black hole mass that is an order of magnitude higher than inferred from the optical lines, X-ray excess variance, and the object classification as a NLS1. Also, the lag model requires a high disk inclination, predicting blueshifted reflection features that are not consistent with the time-averaged spectrum (Figure \ref{fig:consist}).

\subsection{Continuum Lags}
Continuum lags are commonly observed in Galactic black holes \citep{1989Natur.342..773M,1999ApJ...510..874N}, and observed at low frequencies in AGN \citep{2001ApJ...554L.133P, 2004MNRAS.348..783M, 2007ApJ...656..116M,2008MNRAS.388..211A}. The data presented here extend these measurements beyond 10 keV using \nustar data, and show that it is a crucial part of the observed lags, accounting for most of the variance in the data. This highlights the importance of this component, and that it should be explicitly modeled before any relativistic reverberation measurement.

These continuum lags have been attributed to the propagation of accretion fluctuations that modulate large soft-emitting regions before reaching hard-emitting regions, leading to a hard lag \citep{1997MNRAS.292..679L,2001MNRAS.327..799K,2006MNRAS.367..801A}. Although the model we employed is purely mathematical, and any physical model that produces a log-linear lag dependence on energy may be considered, observations of how the variability power changes with energy observed in other sources favors the propagation of fluctuations model \citep[e.g.][]{2014A&ARv..22...72U}.

The data also suggest that these continuum lags scale with the inverse of frequency ($\beta=1$ in the model in equation \ref{eqn:loglin}). This is consistent with the predictions from the standard propagation of fluctuations model \citep{2006MNRAS.367..801A,2017MNRAS.468.3568E}, where the propagation velocity of the flow is tied to the local variability fluctuation time scale. In other words, if typical fluctuations in accretion rate at location $R$ in the disk happen on the viscous time scale $t_\nu$, and that these propagate inward at the local drift velocity $R/t_\nu$, then $\beta=1$ is produced, and the lag is around $1-10\%$ of the fluctuations time scale. The lag scales we measure ($\alpha$ in Table \ref{tab:fit_e1}) in the best fitting models are $331\pm50$ seconds for \mcg and $784\pm86$, $365\pm62$ seconds for \sw, depending on whether a relativistic reflection is included or not. These represent $3\%$ and $4-8\%$ of the time scale normalization of $10^{-4}$ Hz, respectively, as expected for the model. 

We find no evidence of variability in the continuum delays between observations (Table \ref{tab:loglinear_e1}). For \mcg, the lag scale of $\alpha=331\pm50$ seconds is also consistent with values measured from \xmm data at energies below 10 keV \citep[$372\pm 108$ seconds; ][]{2017MNRAS.468.3568E}. The lags in the \xmm data for \sw have not been explicitly modeled \citep{2014MNRAS.440.2347M}. The lack of variability in the parameters of the propagating fluctuations (assuming it is the correct model) is an indication that the speed of propagation and the size of the emission region show very little change during the months spanned by the observations. We note that the spectral modeling of the \mcg data showed changes between observations in the parameters of the coronal component, including the cutoff energy and the inferred optical depth, but no changes in the inner radius of the disk or its emissivity profile \citepalias{2017ApJ...836....2Z}, which may be a proxy for the size of the corona \citep{2012MNRAS.424.1284W}.

\subsection{Relativistic Reverberation}
The \nustar lag data for \mcg we presented show no evidence for relativistic reverberation delays. The continuum model explains all the frequency and energy-dependent delays. The previous report of such a delay did not account for the continuum lags. The data also show that without the continuum lags, a lamp-post relativistic model that is consistent with the time-averaged spectrum is ruled out with high confidence. Once a continuum model is added, it describes all the data. Our approach considers the contribution of the whole relativistic model, which includes both the line at 6.4 keV and the Compton hump.

\begin{figure}
    \centering
    \includegraphics[height=170pt]{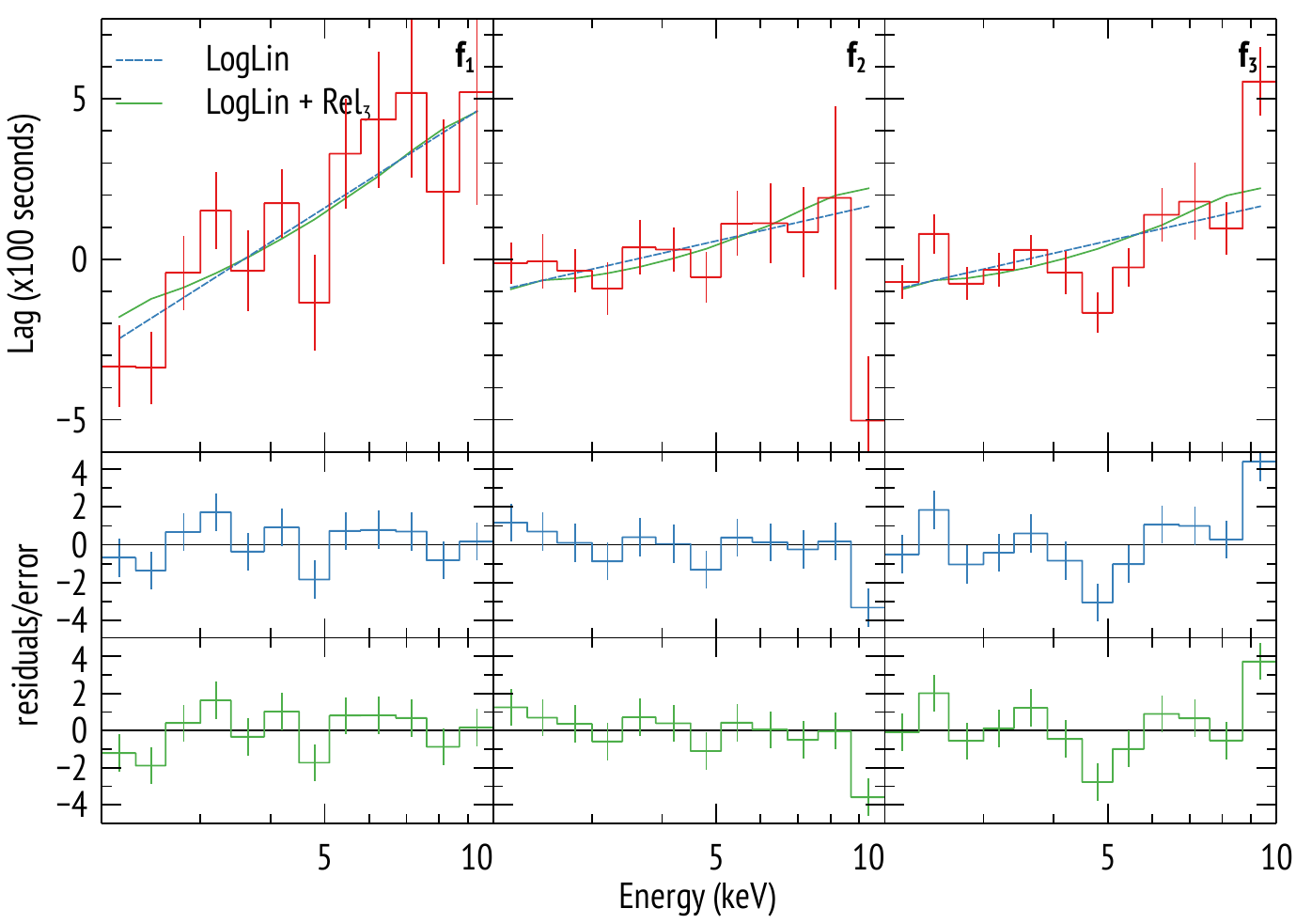}
    \caption{Energy-dependent time lags for \sw from the \xmm data using all archival observations discussed in section \ref{sec:time-avg-fit}. The frequency bins are $10^{-5}, 8\times10^{-5}, 1.5\times10^{-4}, 3\times10^{-4}$ Hz, similar to Figure \ref{fig:lag_e1}. The top row shows the data along with two models from section \ref{sec:other_freq}: a log-linear continuum model with ({\tt LogLin + Rel$_3$}) and without ({\tt LogLin}) relativistic reverberation. The three frequencies in each case are modeled simultaneously as discussed in section \ref{sec:other_freq}. The fit statistics are: $\chi^2/{\rm d.o.f} = 66/34 \, (p=8.6\times10^{-3})$ and $62.5/30\, (p=4.6\times 10^{-3}$ for the {\tt LogLin} and {\tt LogLin + Rel$_3$} models respectively.}
    \label{fig:xmm_sw}
\end{figure}

The \xmm lag data for both sources which show structure around the \FeK line (\citetalias{2014ApJ...789...56Z}, \citealt{2015MNRAS.446..737K}), have not been modeled explicitly with a relativistic reverberation model before, so we employ models similar to those used for the \nustar data. We found that a continuum-only model ({\tt LogLin}) cannot explain the lags. However, adding a relativistic reverberation model does not provide any significant improvement to the fit, regardless of whether the model is forced to be consistent with the time-averaged spectrum or not (i.e. one with free $\theta$, $R_{\rm in}$, $h$ and $M$, similar to {\tt Rel$_3$} in section \ref{sec:results}). This implies that complexity in the \xmm lag data also cannot be described by a relativistic model from a static lamp-post. As an example, Figure \ref{fig:xmm_sw} shows the result of such modeling for \sw, where all 5 observations from section \ref{sec:time-avg-fit} are used to measure lags at frequencies similar to those in Figure \ref{fig:lag_e1}. Again, the continuum {\tt LogLin} model describes most of the variance in the data, while the addition of a relativistic reverberation model from a static lamp-post cannot explain the remaining residuals. We note that the strongest residuals come from the last energy bin. If ignored, the {\tt LogLin} model cannot be statistically rejected, consistent with our \nustar modeling. We plan to present detailed modeling of this and other \xmm data in future work.

We note that the energy-dependent lags in the \nustar data show some apparent features around the iron line in both sources (e.g. panels {\bf f$_2$} in Figure \ref{fig:lag_e1}). The features are however consistent with noise fluctuations around the continuum, and most importantly, they are not at the energies expected given the time-averaged spectrum. For instance, the feature in \mcg has an energy of $4.4\pm0.2$ keV when modeled with a Gaussian function, which is lower than the energy where the iron line peaks (5--6 keV) as seen in the time-averaged spectrum ($\sim6.4$ keV), so a simple association with the iron line is not possible. The low inclination result obtained in the lag model ({\tt Rel$_3$}) is driven by this feature. Complex lag features were also observed in NGC 4151 \citep{2019ApJ...884...26Z}, which were hypothesized to be produced by the complex absorbing system in that source. Even though \mcg is seen through a high column of gas, it is not as variable as in the case of NGC 4151, so an origin in the absorption is not as clear.

\subsection{Comparison With Other Results}
This is the first time explicit modeling of lags above 10 keV is presented. All previous joint spectral-timing modeling used data below 10 keV.
Joint modeling of the time-averaged and lag data for three sources, Mrk 335, IRAS 13224-3809, and Ark 564 using a lamp-post model \citep{2016MNRAS.460.3076C} revealed that only when ionization gradients are included that the lag data can be modeled. The main discrepancy is caused by the lamp-post model based on the time-averaged spectra not producing large enough delay in the iron line \citep[e.g. Figure 10 in][]{2016MNRAS.460.3076C}, or equivalently, producing an apparent dip at 3 keV \citep[e.g.][]{2013MNRAS.434.1129K}. 

\cite{2020ApJ...893...97Z} showed that the lag profile in NGC 5506 is also inconsistent with a static lamp-post model that fits the time-averaged spectra for the same reason. A similar result has also been reported for Mrk 335 by additional modeling by \cite{2020MNRAS.498.4971M}.
The results presented here are consistent with these reports, and they suggest that a static lamp-post model, often used in spectral modeling of time-averaged data is not consistent with the lag data.

\cite{2016MNRAS.458..200W} showed that the dip at 3 keV can be produced by luminosity fluctuations that slowly propagate upward along a vertically extended corona. \cite{2016MNRAS.460.3076C}, on the other hand, suggest that the feature is possibly produced if the effects of ionization gradients in the disc are taken into account. This requires either the source height to be  $>5 r_g$ or the disk is highly ionized at the innermost part and is colder further out. Other deviations from a static lamp-post geometry may also affect the shape of the lag, such as a disk with a finite thickness \citep{2018ApJ...868..109T}. These models, tend to produce shorter delays because of self-shielding.
Some or all of these effects likely occur in real accretion disks. Their effect on the standard spectral modeling of the time-averaged data is not clear and will be crucial to understand.

\subsection{Caveats}
We used a log-linear model as a baseline model against which the relativistic model is tested. It is possible that such a model is not correct, and that curvature due to the cutoff energy for instance may be needed. This seems to be unlikely given the extensive data in stellar-mass black holes \citep{2001MNRAS.327..799K}. Also, the lags in the propagating fluctuations model are related to the size of the emission region and not the physics of emission (e.g. electron temperature).

The values of the lag at the highest observed energies ($>40$ keV) and their uncertainties are important in constraining the parameters in the lag models. We have specifically studied how the uncertainties at those energies are estimated and their effect on the final results. Running Monte Carlo Markov Chains when calculating the lag reveals that the uncertainties are Gaussian for all the energy bands, except for the last bin (in both sources), where the probability distribution has extended wings. We experimented with repeating the modeling in section \ref{sec:other_freq} using different estimates for the lag value and its uncertainty such as the mean and median and using the standard deviation and the 68\% quantiles. We find that the parameter values change slightly, typically within the 1--2$\sigma$ level, but the conclusions based on goodness of fit and model comparison are robust.

In our analysis, we assumed that most of the parameters of the relativistic reverberation model are fixed at those inferred from the time-averaged data, except for those shown in Table \ref{tab:fit_e1}. It may be possible that other parameters need to be varied too. However, the fact that the models are already over-fitting the data, implies that constraining these parameters requires more data than currently available. We have also made the assumption that the variable and the time-averaged spectra (excluding the distant reflector) have the same shape. If the two are different, then the analysis of the former needs to be included. However, more data will be needed to make stronger statements about the consistency of the lag measurements and the shape of the variable spectrum.

\acknowledgments
This material is based upon work supported by the National Aeronautics and Space Administration under Grant No. 80NSSC18K1604 issued through the NuSTAR GO Program. It made use of data from the NuSTAR mission, a
project led by the California Institute of Technology, managed by the Jet Propulsion Laboratory, and funded by the National Aeronautics and Space Administration. This research has made use of the NuSTAR Data Analysis Software (NuSTARDAS) jointly developed by the ASI Science Data Center (ASDC, Italy) and the California Institute of Technology (USA).

\facilities{\nustar, \xmm}

\software{\\
{\tt fqlag} \url{https://github.com/zoghbi-a/fqlag}\\
{\tt kynxilrev} \url{https://projects.asu.cas.cz/stronggravity/kynreverb}\\
{\tt HEASOFT}
}

\bibliography{main}{}
\bibliographystyle{aasjournal}


\listofchanges

\end{document}